\documentclass[epj]{svjour}

\usepackage{graphics}

\begin{document}
\title{\boldmath Spectator interactions and factorization 
in $B\to\pi l\nu$ decay}

\author{M.~Beneke\inst{1} \and Th.~Feldmann\inst{2}}

\institute{Institut f\"ur Theoretische Physik E, RWTH Aachen, 52056
  Aachen, Germany
\and 
CERN Theory Division, 1211 Geneva 23, Switzerland}

\date{{\tt CERN-TH/2003-202}, {\tt PITHA 03/09}, {\tt hep-ph/0308303}}

\abstract{
  We investigate the factorization of different momentum modes 
  that appear in matrix elements for exclusive $B$\/ meson decays into
  light energetic particles for the specific case of $B\to\pi$
  form factors at large pion recoil. We first integrate out hard modes
  with virtualities of order $m_b^2$ ($m_b$ being the heavy quark
  mass), and then hard-collinear modes with virtualities 
  $m_b \Lambda$ ($\Lambda$ being the strong interaction scale).
  The resulting effective theory contains soft and collinear fields
  with virtualities $\Lambda^2$. We prove a previously
  conjectured factorization formula for $B \to \pi$ form factors in
  the heavy quark limit to all orders in $\alpha_s$, paying particular
  attention to `endpoint singularities' that might have appeared in
  hard spectator interactions.
} 

\authorrunning{M.~Beneke, Th.~Feldmann}
\titlerunning{Spectator interactions and factorization 
in exclusive $B\to\pi l\nu$ decay}

\maketitle

\section{Introduction}

In recent years considerable progress in the 
theoretical description of exclusive $B$ meson decays into light
energetic particles has been achieved.
The `QCD-factorization approach' \cite{Beneke:1999br}
allows for a systematic computation of short-distance
QCD corrections in the heavy quark limit. 
It can also be embedded into the soft-collinear 
effective field theory (SCET) framework 
\cite{Bauer:2000yr,Chay:2002vy,Beneke:2002ph,Hill:2002vw}.
Factorization proofs for this class of $B$ decays are complicated,
because two hard scales, $m_b$ and $\sqrt{m_b\Lambda}$, are 
involved by interactions with the spectators in the $B$
meson. Simpler exceptions are the decays $B \to \gamma \ell\nu$ 
\cite{Descotes-Genon:2002mw,Lunghi:2002ju,Bosch:2003fc} (where 
collinear dynamics is absent at leading power) and 
$B \to D\pi$ \cite{Beneke:2000ry,Bauer:2001cu} (where spectator
interactions are absent), for which factorization proofs to all 
orders in perturbation theory have been given in the heavy quark 
limit.

In this article we focus on the superficially simplest `non-trivial'
case, namely the factorization theorem for $B \to \pi$ form factors:
\begin{eqnarray}
F_i(q^2) &=& C_i(q^2) \, \xi_\pi(q^2) + \phi_B \otimes T_i(q^2) \otimes
\phi_\pi \nonumber \\[0.2em]
&& + \,\mbox{power-suppressed terms}
\label{theorem}
\end{eqnarray}
proposed in \cite{Beneke:2000wa}. Here $\xi_\pi(q^2)$ denotes 
a single form factor independent of the 
Dirac structure $\bar q \Gamma_i b$ of the weak decay current.
It reflects the approximate large-recoil symmetry relations for 
heavy-to-light form factors \cite{Charles:1998dr}. Corrections
to these symmetry relations arise from radiative corrections to the
decay vertex and hard spectator scattering. The 
coefficient functions $C_i(q^2)$ and hard-scattering kernels
$T_i(q^2)$ are calculable perturbatively in $\alpha_s$.
In addition, the factorization formula involves the (non-per\-tur\-ba\-tive,
but process-independent) light-cone distribution amplitudes $\phi_B$
and $\phi_\pi$ of the $B$ meson and pion.\footnote{
The first term in (\ref{theorem}) resembles the well-known
factorization formula for $B \to D$ decays where the form factors
factorize into perturbative coefficient functions and a universal
Isgur--Wise form factor \cite{Isgur:vq}. The second term in (\ref{theorem})
is analogous to the factorization formula for the pion form factor
at large momentum transfer \cite{Efremov:1978rn}.}
The present status of the factorization theorem (\ref{theorem}) 
is as follows.

$\bullet$ 
The short-distance functions $C_i(q^2)$ and $T_i(q^2)$ have been 
determined to order $\alpha_s$ \cite{Beneke:2000wa}, confirming the 
structure of (\ref{theorem}). 

$\bullet$ 
 In SCET the functions $C_i(q^2)$
are interpreted as operator-matching coefficients, which arise from 
integrating out hard modes (virtualities of order
$m_b^2$) \cite{Bauer:2000yr}. The renormalisation group in SCET
can be used to sum large logarithms $\ln m_b$.

$\bullet$ 
The hard-scattering kernels $T_i(q^2)$ are obtained 
by integrating out hard-collinear modes (see Table~\ref{table1} below)
that arise from interactions of the external
soft and colline\-ar fields, and have virtualities of
order $m_b\Lambda$. In this way the tree-level expressions 
for $T_i(q^2)$ have been recovered \cite{Bauer:2002aj}, and the
general procedure for a two-step matching procedure in SCET to all orders
in perturbation theory has been outlined. 

A proof of (\ref{theorem}) to all orders in
$\alpha_s$ requires a detailed analysis of the second matching step, 
in which hard-col\-linear modes are integrated out perturbatively, since
only then can the problem of convergence of the soft and col\-line\-ar
convolution integrals in (\ref{theorem}) be addressed. 
Possible `endpoint divergences' are related to the fact that 
the resulting operators in the effective theory do
not necessarily factorize into purely soft or collinear hadronic
quantities in the conventional sense. To prove (\ref{theorem}), 
we have to define $\xi_\pi(q^2)$, and show that the remainder 
factorizes into conventional distribution amplitudes convoluted 
with a hard-scattering kernel $T_i(q^2)$, free of 
endpoint singularities. 

\begin{figure}[t]
\begin{center}
\resizebox{0.3\textwidth}{!}{%
  \includegraphics{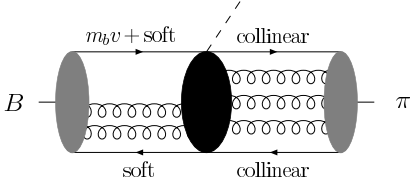}}
\end{center}
\caption{Kinematics for $B \to \pi$ form factor at large recoil.}
\label{fig:1}   
\end{figure}

In the following we sketch the main steps that lead to the completion 
of the factorization proof (for details see \cite{Beneke:2003}). 
The effective theory set-up is similar 
to \cite{Bauer:2002aj}: we perform the matching in two steps, onto 
an effective theory in which the degrees of freedom are quark and
gluon fields with virtualities of order $\Lambda^2$.
In the rest frame of the $B$ meson, the initial state consists of only
soft modes with all momentum components of order $\Lambda$.
The final state pion is made up only of collinear 
modes (Fig.~\ref{fig:1}). 
Hard and hard-collinear modes have been integrated out and are 
encoded as perturbative
coefficient functions multiplying effective operators.
Table~\ref{table1} summarizes the various modes and the scaling
of the momentum components defined 
with respect to two light-cone vectors $n_\pm$ (the pion momentum
is $p_\pi = E n_-$, and for the heavy quark velocity we take 
$v_\perp = 0$). 
The expansion parameter in the effective theory is  
$\lambda = \sqrt{\Lambda/m_b}$.\footnote{The only
objects scaling with an odd power of $\lambda$ are
soft quark fields and meson states. Because $B$ mesons  
contain an even number of soft quark fields, the expansion parameter
for exclusive matrix elements in the effective theory is in fact 
$\lambda^2 = \Lambda/m_b$, but the absolute scaling may be an odd
power of $\lambda$.}

\begin{table}[hbt]
\caption{Scaling of momentum modes relevant to $B \to \pi$ form factors. 
The entries in brackets denote an alternative terminology, which can be
found in the literature. }
\label{table1}     
\begin{center}
\begin{tabular}{lll}
\hline\noalign{\smallskip}
$(n_+p,p_\perp,n_-p)$ & terminology & (alternative)  \\
\noalign{\smallskip}\hline\noalign{\smallskip}
$(1,1,1)$ & hard & (hard) \\
$(1,\lambda,\lambda^2)$ & hard-collinear & (collinear) \\
\noalign{\smallskip}\hline
&&\\[-0.3cm]
$(\lambda^2,\lambda^2,\lambda^2)$ & soft & (ultrasoft) \\
$(1,\lambda^2,\lambda^4)$ & collinear & (ultracollinear) \\
\noalign{\smallskip}\hline
\end{tabular}
\end{center}
\end{table}

\section{Results}

We first checked the relevance of the different modes in
Table~\ref{table1} 
by considering a toy loop integral with the same
kinematics as in $B \to \pi$. The expansion of the integral is reproduced  
by momentum regions along the lines of \cite{Beneke:1997zp}, including
separate hard-collinear and collinear regions.
The collinear integral suffers from an endpoint di\-ver\-gence 
(not regularized in $d$ dimensions),
when the $n_+ k$ component of the loop momentum becomes small.
This divergence is cancelled by an endpoint divergence in the 
soft region, arising from $n_- k \to 0$. The relation between
endpoint divergences in soft and collinear integrals prevents 
the naive factorization of the form factor 
into light-cone distribution amplitudes 
(in which case we could have set $\xi_\pi(q^2)$  
in (\ref{theorem}) to zero).

Turning to the construction of the effective theory, 
we first integrate out hard modes at a
factorization scale $\mu\sim m_b$, and arrive at an effective theory 
SCET(hc,c,s), where soft and collinear fields interact 
by exchange of hard-collinear modes (heavy quarks only interact with
soft modes). The first step consists of classifying all current 
operators in SCET(hc,c,s), which contribute to the leading-power form
factor. To this end we determine the additional $\lambda$ suppression
factors in the matrix elements of SCET(hc,c,s) operators incurred by
the necessity to convert hard-collinear fields into collinear and 
soft fields via power-suppressed interaction terms. 
We then find only two possible structures 
\cite{Beneke:2002ph,Pirjol:2002km}:
\begin{eqnarray}
&& \tilde C_{ij}\!\left(\frac{{\cal E}}{m_b},\frac{\mu}{m_b}\right)\,
  \bar \xi_{\rm hc}(0)\Gamma_j h_v(0),
\label{symm}
\\[0.2em]
&& \frac{1}{m_b} \,
 \int d\hat{s} \;
 \tilde C_{ij\perp}'\!\left(\hat{s}, \frac{{\cal E}}{m_b},
 \frac{\mu}{m_b}\right)\bar \xi_{\rm hc/c}(0) A^\perp_{\rm hc}(s n_+) 
    \Gamma_j h_v(0),
\cr &&
\label{break}
\end{eqnarray} 
where ${\cal E}=(n_- v)(n_+ P)/2$, and $P$ is the (collinear) momentum 
operator. In the second line the quark field can be collinear or 
hard-collinear. We also adopt the gauge $n_+ A_{\rm hc/c}=n_- A_s=0$. The
gauge transformations that undo light-cone gauge introduce collinear
and soft Wilson lines into the operators. The projection properties of
the $\xi$ and $h_v$ fields imply that
$\Gamma_j=\{1,\gamma_5,\gamma_\perp^\mu\}$ is the most general Dirac
structure. The short-distance coefficients from integrating out 
hard modes may depend only on the boost-invariant and dimensionless
ratios ${\cal E}/m_b$, $\mu/m_b$ and $\hat{s} = m_b s/(n_-v)$. 
Note that in SCET(hc,c,s) the operator (\ref{break}) is suppressed  
relative to (\ref{symm}), but both contribute to the leading-power 
form factor \cite{Bauer:2002aj}.

We now define the $C_i(q^2)\xi_\pi(q^2)$ term in (\ref{theorem}) 
as the matrix element of the operator (\ref{symm}). Since only 
$\Gamma_j=1$ does not vanish between $\langle\pi|$ and $|B\rangle$, 
this defines one non-perturbative form 
factor independent of the original Dirac
structure. It remains to show that the second operator  
(\ref{break}) factorizes into light-cone distribution
amplitudes (LCDAs). Although we do not need to further discuss the operator
(\ref{symm}) at this point, we note that if one 
attempted to factorize it into LCDAs, the resulting convolution 
integrals would be divergent. Furthermore, three-particle LCDAs of the
$B$ meson and pion would appear even at leading power (as also 
noted in \cite{Lange}). We also find that 
the matrix element of (\ref{symm}) scales 
as $\lambda^3$, which confirms the well-known scaling 
of heavy-to-light form factors at large recoil.

In the second matching step hard-collinear modes are
integrated out at $\mu \sim \sqrt{m_b \Lambda}$. The resulting effective
theory is denoted by SCET(c,s). The following results and arguments
are relevant to the factorization proof:

$\bullet$ 
Soft and collinear fields are decoupled  
in the relevant terms of the SCET(c,s) Lagrangian.

$\bullet$ 
Two-quark operators in SCET(c,s) do not contribute 
to the hard-scattering amplitudes $T_i(q^2)$, because the 
col\-linear (soft) 
fields in the operator must have the same quantum numbers as 
the pion ($B$ meson). The relevant terms in the matching 
of (\ref{break}) on SCET(c,s) operators must therefore 
contain the fields $[\bar\xi_c\xi_c][\bar q_s h_v]$, and possibly 
additional gluon or quark fields with the correct 
quantum numbers.

$\bullet$ 
Dimensional analysis and boost invariance imply that the operator 
(\ref{break}) matches only onto four quark operators at leading power,  
\begin{eqnarray}
&&   \bar \xi_{\rm hc/c}(0) A^\perp_{\rm hc}(s n_+) 
    \Gamma_j \,  h_v(0)
\nonumber \\[0.3em]
&&\to 
\int ds' dt  \, \tilde C_{jk}''\!
\left(\hat{s};\ln [\mu^2 s't], n_+ P s'\right)
 \nonumber \\[0.1em]
&& 
\hspace*{0.8cm} \times \, [\bar \xi_c(s'n_+) \Gamma_k h_v(0)]
\, [\bar q_s(t n_-) \Gamma_k\xi_c(0)] .
\label{tcontrib}
\end{eqnarray}
The right-hand side scales as $\lambda^8$, implying a contribution
to the form factor at order $\lambda^3$, which is as large as the 
$C_i(q^2)\xi_\pi(q^2)$ term. We also note that the effective current 
(\ref{tcontrib}) must be local in the transverse direction. 

$\bullet$ 
The right-hand side of (\ref{tcontrib}) has the required structure 
$\phi_B \otimes T_i(q^2) \otimes\phi_\pi$ after taking the $B\to\pi$
matrix element, but it remains to show that
the convolution integrals converge (under the assumption of the
standard endpoint behaviour of the LCDAs). Boost invariance implies
that the short-distance coefficients $\tilde C_{jk}''$
may depend only logarithmically on the light-cone distance
$t$ between the heavy quark and the soft spectator, and that 
only the plus-projection of the $B$ meson light-cone matrix element
appears in (\ref{tcontrib}). After taking the $B\to\pi$ matrix
element, the $t$-convolution integral is 
therefore 
\begin{eqnarray}
  \int dt \, \tilde C_{jk}''(\ln t) \, \tilde \phi_+^B(t). 
\end{eqnarray}
This integral converges, given the endpoint behaviour of
the $B$ meson light-cone distribution amplitude $\tilde \phi_+^B(t)$ 
\cite{Lange:2003ff}.

$\bullet$ 
It follows that the collinear convolution integrals over $\hat s$ and 
$s'$ also converge. If this were not the case, the collinear
convolution integral would have to be regulated. The regulator
dependence introduced by the endpoint divergence would have to cancel
against a soft endpoint divergence (since the effective theory is
assumed to reproduce the infrared physics correctly). Since there is
none, the collinear convolution integrals must be finite. 

This completes the proof of the factorization formula 
(\ref{theorem}). Both terms in the formula are of order $\lambda^3$, 
as far as heavy quark power counting is concerned. The hard
scattering term is proportional to $\alpha_s(\sqrt{m_b\Lambda})$. 
It is an open question whether the $C_i(q^2)\xi_\pi(q^2)$ term 
is proportional to 1 (as assumed in 
\cite{Bauer:2000yr,Beneke:2000wa,Charles:1998dr}), or 
proportional to $\alpha_s(\sqrt{m_b\Lambda})$. The clarification of
this point requires an analysis of large logarithms  
in SCET(c,s).

\section{Summary and outlook}

We have proved the factorization formula (\ref{theorem}) for $B \to \pi$
transition form factors in the heavy quark limit. 
Attempting 
to factorize heavy-to-light form factors along the same lines as 
the $\pi\to\pi$ transition form factor \cite{Efremov:1978rn}, 
it is found that three-particle Fock states of the $B$ meson and the
pion contribute at leading power, and that the convolution 
integrals are in general ill-defined. We have shown that all 
this is part of a single form factor
$\xi_\pi(q^2)$, while the remaining contributions can be factorized 
in the standard way. This confirms the factorization conjecture of 
\cite{Beneke:2000wa} and completes the factorization argument of 
\cite{Bauer:2002aj}. Our result generalizes to other heavy-to-light
form factors ($B \to \rho$ etc.). We remark that
in the case of the $B \to \gamma$ form factor, the endpoint divergences and
`non-factorizable' contributions only appear at subleading order
in the heavy quark expansion, when the hadronic structure of the photon
is resolved. Our framework can also be adapted to situations
with two collinear directions, such as for the pion electromagnetic
form factor at large momentum transfer. For details 
we refer to a forthcoming paper \cite{Beneke:2003}.

\subsection*{Acknowledgements}

We would like to thank S.~Bosch, S.~Descotes--Genon,
B.~Lange, E.~Lun\-ghi, D.~Pirjol, and I.~Stewart
for interesting discussions.

\end{document}